\documentclass[preprint,12pt]{elsarticle}

\usepackage[utf8]{inputenc}
\usepackage{amsmath}
\usepackage{amsfonts}
\usepackage{amssymb}
\usepackage{graphicx}
\usepackage{booktabs}

\newtheorem{defn}{Definition}

\newtheorem{rem}[defn]{Remark}
\newtheorem{thm}[defn]{Theorem}

\newtheorem{lem}[defn]{Lemma}

\begin{document}

\begin{frontmatter}

\title{Exponential Utility Maximization with Delay in a Continuous Time Gaussian Framework}

\author{Yan Dolinsky}

\affiliation{organization={Department of Statistics, Hebrew University},
          city={Jerusalem},
          country={Israel}}

\begin{abstract}
In this work we study the continuous time exponential utility maximization
problem in the framework of an
investor who is informed about
the risky asset’s price changes with a delay. This leads to a non-Markovian stochastic control problem.
In the case where the risky asset
is given by a Gaussian process (with some additional properties) we
establish a solution for the optimal control and the corresponding value.
Our approach is purely probabilistic
 and is based on the
theory for Radon-Nikodym derivatives of Gaussian measures
developed by Shepp \cite{S:66} and 
Hitsuda \cite{H:68}. 
\end{abstract}


\end{frontmatter}

\section{Introduction and the Main Result}\label{sec:1}

Taking into account frictions is an important challenge in financial modelling.
In this note, we focus on the friction arising from the fact
that investment decisions may be based only on delayed information,
and the actual present market price is unknown at the time of decision making.
This corresponds to the case where there is a time
delay in receiving market information (or in applying it), which causes the trader’s
filtration to be delayed with respect to the price filtration.

We consider a simple financial market consisting of one risky asset. We assume that the law of the risky asset is a
Gaussian measure which is equivalent to the Wiener measure.
Roughly speaking, this property can be achieved by a linear transformation of the risky asset, provided that
the initial law of the risky asset is a Gaussian measure which is equivalent to a continuous Gaussian martingale.
We allow a general (deterministic) delay function and our main result Theorem \ref{thm1}
can be viewed as an extension of the results from \cite{DZ}
to the continuous time setup and a general (not necessarily constant) delay.

Next, we describe the setting and notation.
Let $T>0$ be a fixed time horizon and let $C$ be the space of all continuous function
$\phi:[0,T]\rightarrow\mathbb R$ equipped with the uniform norm. Denote by $\mathcal B$ the corresponding Borel $\sigma$-algebra. The canonical process
$X=(X_t)_{0\leq t\leq T}$ is given by
$X_t(\omega)=\omega(t)$, $\omega\in C$, $t\in [0,T]$.
By $\mathbb W$ we denote the Wiener measure on $(C, \mathcal B)$.
We call a probability measure $\mathbb Q$ on $(C,\mathcal B)$ a Gaussian measure if the canonical process $X$ is a Gaussian process
with respect to $\mathbb Q$.
Such a measure determined by its mean
$\mathbb E_{\mathbb Q}[X_t]$, $t\in [0,T]$ and its covariance
$\mathbb E_{\mathbb Q}\left[\left(X_t-\mathbb E_{\mathbb Q}[X_t]\right)\left(X_s-\mathbb E_{\mathbb Q}[X_s]\right)\right]$, $t,s\in [0,T].$

Any function $\psi\in L^2([0,T]^2)$ induces
a Hilbert-Schmidt operator $\Psi:L^2[0,T]\rightarrow L^2[0,T]$ given by
$\left(\Psi(\varphi)\right)(t)=\int_{0}^T \psi(t,s) \varphi(s) ds$, $\varphi\in L^2[0,T]$, $t\in [0,T]$.
From the classical theory of Hilbert-Schmidt operators (see \cite{S:58}) the spectrum $Spec(\Psi)$
consists of at most countable many points and every non-zero value in $Spec(\Psi)$
is an eigenvalue of finite multiplicity. If $\psi$ is symmetric i.e. $\psi(t,s)=\psi(s,t)$ a.s. then
the corresponding
 Hilbert-Schmidt operator is self adjoint and so $Spec(\Psi)\subset\mathbb R$. Let $\mathcal S\subset L^2([0,T]^2)$ be the set of all
 symmetric functions such that the spectrum of the corresponding Hilbert-Schmidt operator
 is a subset of $(-\infty,1)$.

Consider a financial market with one risky asset given by the canonical process
$X=(X_t)_{0\leq t\leq T}$ and a savings account that,
for simplicity, bears no interest.
Let $\mathbb P$ be the market probability measure, this is a probability measure on the space
$(C,\mathcal B)$. Assume that $\mathbb P\sim\mathbb W$ is a Gaussian probability measure
which is equivalent to the Wiener measure $\mathbb W$.

The following result from \cite{C:03} characterizes Gaussian probability measures being equivalent to the Wiener measure and plays
the key input, together with the theory for Radon-Nikodym derivatives of
Gaussian measures developed by Shepp \cite{S:66} and Hitsuda \cite{H:68}, to prove our main
result. From Theorems 1-2 in \cite{C:03}
$\mathbb P\sim\mathbb W$ if and only if
there exist $\tilde a\in L^2[0,T]$
and $\tilde f\in \mathcal S$ such that
\begin{align}\label{shepp1}
&\mathbb E_{\mathbb P}[X_t]=\int_{0}^t \tilde a(s)ds, \ \ t\in [0,T], \ \ \mbox{and}\\
&\mathbb E_{\mathbb P}\left[\left(X_t-\mathbb E_{\mathbb P}[X_t]\right)\left(X_s-\mathbb E_{\mathbb P}[X_s]\right)\right]\nonumber\\
&=
\min(t,s)-\int_{0}^t\int_{0}^s \tilde f(u,v)dudv, \ \ t,s\in [0,T].\label{shepp2}
\end{align}

From the general theory of integral equation (see \cite{S:58}) it follows that there exists a unique element $f\in \mathcal S$ such that
\begin{equation}\label{shepp3}
f(t,s)+\tilde f(t,s)=\int_{0}^T f(t,u)\tilde f(u,s)du,  \ \ t,s\in [0,T].
\end{equation}
Let
\begin{equation}\label{shepp4}
a(t):=\tilde a(t)-\int_{0}^T f(t,s)\tilde a(s) ds, \ \ t\in [0,T],
\end{equation}
and
\begin{equation}\label{shepp5}
c:=\frac{1}{2}\left(\sum_{i=1}^N \left(\lambda_i +\log(1-\lambda_i)\right)-\int_{0}^T a_t \tilde a_t dt\right)
\end{equation}
where $\lambda_j$, $j=1,...,N$, $N\in\mathbb N\cup\{\infty\}$
are the non-zero eigenvalues
of the Hilbert-Schmidt operator induced by $f$, repeated according to their multiplicity.
From Theorem 1 in \cite{C:03} the corresponding Radon-Nikodym derivative is given by
\begin{equation}\label{1}
\frac{d\mathbb P}{d\mathbb W}= \exp\left(c+\int_{0}^T a(s)dX_s+\int_{0}^T\int_{0}^s f(s,u)dX_udX_s\right)
\end{equation}
where the above integrals are  It\^{o} integrals.

Let $\mathcal F_t$, $0\leq t\leq T$ be the augmented (with respect to $\mathbb P$) filtration which
is generated by $X_t$, $0\leq t\leq T$.
Let $\tau:[0,T]\rightarrow [0,T]$ be a nondecreasing and right-continuous function.
The investor's flow of information is given by the (right-continuous and complete) filtration
$\mathcal G_t:=\mathcal F_{\tau(t)}$, $t\in [0,T]$.
We assume that there exists $\epsilon>0$ such that
$\tau(t)\leq (t-\epsilon)^{+}$ for all $t$.
This inequality can be viewed as an assumption
of strict delay during all time period.
Let $\tau^{-1}(t):=T\wedge\inf\left\{s\in [0,T] \ : \ \tau(s)\geq t \right\}$, $t\in [0,T]$ denote the left-continuous
inverse of $\tau$.

A trading strategy is a progressively measurable
process $\gamma=\left(\gamma_t\right)_{0\leq t\leq T}$
with respect
to $(\mathcal G_t)_{t\in [0,T]}$ which satisfies $\int_{0}^T \gamma^2_t dt<\infty$ a.s.
The corresponding portfolio value
at the maturity date is given by
$\int_{0}^T \gamma_t dX_t$.
Let us emphasize that we do not require any notion of admissibility for
the portfolio value.
Denote by $\mathcal A$ the set of all trading strategies.

The investor’s preferences are described by an exponential utility function
$u(x)=-\exp(-\alpha x)$, $x\in\mathbb R$,
 with absolute risk aversion parameter $\alpha>0$, and her goal is to
\begin{equation}\label{2}
\mbox{Maximize} \ \ \mathbb E_{\mathbb P}\left[-\exp\left(-\alpha \int_{0}^T \gamma_t dX_t\right)\right] \ \ \mbox{over} \ \ \gamma\in\mathcal A.
\end{equation}
Without loss of generality, we take the risk aversion $\alpha=1$.

We arrive at the main result of the paper.
Let $\mathcal V$ be the set of all functions
$\phi\in L^2([0,T]^2)$ such that $\phi(t,s)=0$ if $s>t$ (Volterra Kernels).
\begin{thm}\label{thm1}
There exists a unique pair $(\kappa,g)\in\mathcal V\times\mathcal S$ such that
$\kappa(t,s)=0$ for $t<\tau^{-1}(s)$, $g(t,s)=0$ for $t\geq \tau^{-1}(s)$ and
\begin{equation}\label{3}
 f(t,s)-\kappa(t,s)+g(t,s)=\int_{s}^T \left(f(t,u)-\kappa(t,u)\right) g(u,s) du, \ \ 0\leq s\leq t\leq T.
\end{equation}
The maximizer $\hat\gamma=(\hat\gamma_t)_{0\leq t \leq T}$ for the optimization problem (\ref{2}) is unique and is given by the linear form
\begin{equation}\label{4}
\hat\gamma_t=a(t)+\int_{0}^t \kappa(t,s) dX_s.
\end{equation}
The corresponding value is given by
\begin{align}\label{5}
&\mathbb E_{\mathbb P}\left[-\exp\left(-\int_{0}^T\hat\gamma_t dX_t \right)\right]\nonumber\\
&=
-\exp\left(c-\int_{0}^T\int_{0}^s f(s,u)\tilde g(s,u)du ds-\frac{1}{2}\int_{0}^T\int_{0}^s  g^2(s,u)du ds\right)
\end{align}
where $\tilde g(s,u):=g(s,u)-\int_{0}^u g(s,v)g(u,v)dv$, $0\leq u\leq s$.
\end{thm}
\begin{rem}
Observe that for any 
$\alpha>0$ we have 
$\alpha \mathcal A=\mathcal A$. Hence, the value of the exponential utility maximization problem does not depend on the risk aversion $\alpha$. 
Moreover, for a general $\alpha>0$,  the strategy $\frac{1}{\alpha}\hat\gamma$ (where $\hat\gamma$ is given by (\ref{4})) is the (unique) maximizer for 
(\ref{2}).
\end{rem}

\subsection{Computational Example}
Let $T=1$ be the maturity date and let $\mathbb P$ be the
distribution of the
Gaussian process $(B_t+t Z)_{0\leq t\leq 1}$
where $B$ is a Brownian motion and $Z\sim N(\mu,\sigma^2)$ is a normal random variable (with mean $\mu$ and variance $\sigma^2$)
which is independent of $B$.
Thus, (\ref{shepp1})-(\ref{shepp2}) hold true for the constant functions
$\tilde a\equiv \mu$ and $\tilde f\equiv -\sigma^2$.
In this case equation (\ref{shepp3}) has the constant solution
$f\equiv \frac{\sigma^2}{1+\sigma^2}$. From (\ref{shepp4}) we obtain
$a\equiv \frac{\mu}{1+\sigma^2}$. Clearly,
for the Hilbert-Schmidt operator which corresponds to the constant function $f\equiv \frac{\sigma^2}{1+\sigma^2}$
 there exists a unique non-zero eigenvalue $\lambda=\frac{\sigma^2}{1+\sigma^2}$.
Hence,
(\ref{shepp5}) gives
$c=\frac{\sigma^2-\mu^2}{2(1+\sigma^2)}-\frac{1}{2}\log\left(1+\sigma^2\right)$.

Next,
from (\ref{3})
and the fact that $\kappa(t,s)=0$ for $t<\tau^{-1}(s)$
it follows that for
$t\in [s,\tau^{-1}(s))$ we have the relation
$\frac{\sigma^2}{1+\sigma^2}+g(t,s)=\frac{\sigma^2}{1+\sigma^2}\int_{s}^{\tau^{-1}(s)}g(u,s)du$.
 Hence,
 \begin{equation}\label{g}
g(t,s)=-\mathbb I_{t<\tau^{-1}(s)} \frac{\sigma^2}{1+\sigma^2\left(1+s-\tau^{-1}(s)\right)}, \ \ 0\leq s\leq t\leq 1.
\end{equation}
The Fubini theorem yields
\begin{equation*}
\int_{0}^1\int_{0}^s  g^2(s,u)du ds=\int_{0}^1  \left(\frac{\sigma^2}{1+\sigma^2\left(1+u-\tau^{-1}(u)\right)}\right)^2 (\tau^{-1}(u)-u) du
\end{equation*}
and
\begin{align*}
&\int_{0}^1\int_{0}^s f(s,u)\tilde g(s,u)du ds\\
&=\frac{\sigma^2}{1+\sigma^2}\int_{0}^1\int_{0}^s\left(g(s,u)-\int_{0}^u g(s,v)g(u,v)dv\right)duds\\
&=-\int_{0}^1\frac{\sigma^4\left(\tau^{-1}(u)-u\right)}{(1+\sigma^2)\left(1+\sigma^2\left(1+u-\tau^{-1}(u)\right)\right)}du\\
&-\frac{1}{2}\frac{\sigma^2}{1+\sigma^2}\int_{0}^1\left(\frac{\sigma^2 \left(\tau^{-1}(v)-v\right)}{1+\sigma^2\left(1+v-\tau^{-1}(v)\right)}\right)^2dv.
\end{align*}
From (\ref{5}) we conclude that the value of the optimization problem (\ref{2}) is given by
$$-\frac{1}{\sqrt{1+\sigma^2}}\exp\left(\frac{\sigma^2-\mu^2}{2(1+\sigma^2)}+\frac{\sigma^4}{2(1+\sigma^2)}\int_{0}^T \frac{\tau^{-1}(t)-t}
{1+\sigma^2\left(1+t-\tau^{-1}(t)\right)}dt\right).$$
We can view the term $\frac{\sigma^4}{2(1+\sigma^2)}\int_{0}^T \frac{\tau^{-1}(t)-t}
{1+\sigma^2\left(1+t-\tau^{-1}(t)\right)}dt$ as the penalty (on a logarithmic scale)
for the delay in the information flow.

Next, we treat the optimal portfolio. Observe that
$t\geq\tau^{-1}(s)\Leftrightarrow \tau(t)\geq s$. Thus,
from (\ref{3}) and (\ref{g})
we obtain that $\kappa(t,s)$, $0\leq s\leq \tau(t)$
is the unique solution
to the
Volterra integral equation
$$\kappa(t,s)=\frac{\sigma^2}{1+\sigma^2\left(1+s-\tau^{-1}(s)\right)}\left(1-\int_{s}^{\tau^{-1}(s)}\kappa(t,u)du\right).
$$
From the equality
$$\mathbb E_{\mathbb P}\left[\left(X_t-\mu_t\right)\left(X_s-\mu s\right)\right]=
\min(t,s)\left(1+\sigma^2 \max(t,s)\right), \ \ t,s\in [0,T]$$ we conclude that
$\mathbb P$ is a distribution of a Gauss-Markov process (see \cite{B}).
Still, due to the delay, the stochastic control problem (\ref{2})
is not Markovian.
In view of Theorem \ref{thm1} we obtain that the optimal control is a sum of the
constant $\frac{\mu}{1+\sigma^2}$
and a Gaussian-Volterra process with a Kernel which solves the above
Volterra integral equation.

\section{Proof of Theorem \ref{thm1}}\label{sec:2}
\begin{lem}\label{lem2}
Let $h\in \mathcal S$ such that
$h(t,s)=0$ for $t\geq\tau^{-1}(s)$. Then,
for any $\gamma\in\mathcal A$
\begin{align*}
&\mathbb E_{\mathbb P}\left[-\exp\left(-\int_{0}^T \gamma_t dX_t\right)\right]\leq\\
&-\exp\left(c-\int_{0}^T\int_{0}^s f(s,u)\tilde h(s,u)du ds-\frac{1}{2}\int_{0}^T\int_{0}^s  h^2(s,u)du ds\right)
\end{align*}
where $\tilde h(s,u):=h(s,u)-\int_{0}^u h(s,v)h(u,v)dv$, $0\leq u\leq s$.
\end{lem}
\textbf{Proof:}
Let $\gamma\in\mathcal A$. Without loss of generality we assume that (otherwise the statement is trivial)
\begin{equation}\label{6+}
\mathbb E_{\mathbb P}\left[\exp\left(-\int_{0}^T \gamma_t dX_t\right)\right]<\infty.
\end{equation}
Standard density arguments yield that there exists a sequence $h_n\in\mathcal V$, $n\in\mathbb N$
which converge to $h$ in $L^2([0,T]^2)$ such that for all $n$,
$h_n(t,s)=0$ for $t>\tau^{-1}\left((s-\frac{1}{n})^{+}\right)$.
In order to prove the statement of the lemma it is sufficient to establish that for any $n\in\mathbb N$
\begin{align}\label{7}
&\mathbb E_{\mathbb P}\left[\exp\left(-\int_{0}^T\gamma_t dX_t \right)\right]\nonumber\\
&\geq
\exp\left(c-\int_{0}^T\int_{0}^s f(s,u)\tilde h_n(s,u)du ds-\frac{1}{2}\int_{0}^T\int_{0}^s  h^2_n(s,u)du ds\right)
\end{align}
where $\tilde h_n(s,u):=h_n(s,u)-\int_{0}^u h_n(s,v)h_n(u,v)dv$, $0\leq u\leq s$.

Fix $n\in \mathbb N$. Since $h_n\in\mathcal V$ then (see Theorem 2 in \cite{C:03})
there exists a probability measure
$\mathbb Q_n\sim\mathbb P$ and a
$\mathbb Q_n$-Brownian motion $(B^n_t)_{0\leq t\leq T}$
such that the canonical process satisfies
$X_t=B^n_t-\int_{0}^t \int_{0}^s h_n(s,u)dB^n_u ds$, $t\in [0,T]$.
From the Girsanov theorem and the It\^{o} Isometry we have
\begin{equation}\label{7+}
\mathbb E_{\mathbb Q_n}\left[\log\left(\frac{d\mathbb Q_n}{d\mathbb W}\right)\right]=\frac{1}{2}\int_{0}^{T}\int_{0}^s h^2_n(s,u)du ds.
\end{equation}
From the Fubini theorem and the It\^{o} formula we obtain
\begin{align}\label{7++}
&\mathbb E_{\mathbb Q_n}\left[\int_{0}^T\int_{0}^s f(s,u)dX_udX_s\right]\nonumber\\
&=\mathbb E_{\mathbb Q_n}\left[\int_{0}^T\left(\int_{0}^s\int_{0}^u f(s,u) h_n(u,v)dB^n_v du-\int_{0}^s f(s,v)dB^n_v \right)\times\right.\nonumber\\
&\left. \int_{0}^s h_n(s,v)dB^n_v ds\right]=-\int_{0}^T\int_{0}^s f(s,u)\tilde h_n(s,u)du ds.
\end{align}

Next, set $\Upsilon:=\frac{d\mathbb Q_n}{d\mathbb W}$ and
$$\Xi:=\int_{0}^T\int_{0}^s f(s,u)dX_udX_s+\int_{0}^T a_t dX_t+\left(\int_{0}^T\gamma_tdX_t\right)^{-}.$$
 From (\ref{1}) and (\ref{6+}) we have
$\mathbb E_{\mathbb W}\left[e^{\Xi}\right]<\infty$. Thus,
from the classical Legendre-Fenchel duality inequality
$\Xi \Upsilon \leq e^{\Xi} + \Upsilon\left(\log \Upsilon-1\right)$, (\ref{7+})-(\ref{7++})
and the fact that
$\int_{0}^T a(t)dX_t\in L^1(\mathbb \mathbb Q_n)$
we obtain
\begin{equation}\label{abc}
\mathbb E_{\mathbb Q_n}\left[\left(\int_{0}^T \left(\gamma_t-a(t)\right) dX_t\right)^{-}\right]<\infty.
\end{equation}
From the Fubini theorem
$$\int_{0}^T \left(\gamma_t-a(t)\right) dX_t=\int_{0}^T \left(\gamma_t-a(t)-\int_{t}^T \left(\gamma_u-a(u)\right) h_n(u,t)du\right)dB^n_t.$$
Recall that there exists $\epsilon>0$ such that $\tau(t)\leq (t-\epsilon)^{+}$ for all $t$.
This together with the relation
$h_n(u,t)=0$ for $u>\tau^{-1}\left((t-\frac{1}{n})^{+}\right)$ yields that for sufficiently large $n$
the stochastic process
$\gamma_t-a(t)-\int_{t}^T \left(\gamma_u-a(u)\right) h_n(u,t)du$, $t\in [0,T]$  is progressively measurable
with respect to the filtration $\mathcal F_{\left(t-\frac{1}{n}\right)^{+}}$, $t\in [0,T]$.
Since $B^n$ is a $\mathbb Q_n$-Brownian motion, then from Lemma 5.3 in \cite{DZ1} and (\ref{abc}) we conclude that
$\mathbb E_{\mathbb Q_n}\left[\int_{0}^T (\gamma_t-a(t)) dX_t\right]=0$. Hence, from (\ref{7+})-(\ref{7++})
\begin{align}\label{8}
&\mathbb E_{\mathbb Q_n}\left[\tilde\Xi-\log\left(\frac{d\mathbb Q_n}{d\mathbb W}\right)\right]\nonumber\\
&=-\int_{0}^T\int_{0}^s f(s,u)\tilde h_n(s,u)du ds-
\frac{1}{2}\int_{0}^{T}\int_{0}^s h^2_n(s,u)du ds
\end{align}
where
$\tilde\Xi:=\int_{0}^T\int_{0}^s f(s,u)dX_udX_s-\int_{0}^T\left(\gamma_t-a(t)\right)dX_t$.

Finally, from  the Legendre-Fenchel inequality and (\ref{1}) we get that for
$z:=\exp\left(c+\mathbb E_{\mathbb Q_n}\left[\tilde\Xi-\log\left(\frac{d\mathbb Q_n}{d\mathbb W}\right)\right]\right)$
\begin{align*}
&\mathbb E_{\mathbb P}\left[\exp\left(-\int_{0}^T \gamma_t dX_t\right)\right]=
\mathbb E_{\mathbb W}\left[e^{c+\tilde\Xi}\right]\\
&\geq \mathbb E_{\mathbb W}\left[(c+\tilde\Xi)z \frac{d{\mathbb Q_n}}{d\mathbb W}-
z \frac{d{\mathbb Q_n}}{d\mathbb W}\left(\log\left(z \frac{d{\mathbb Q_n}}{d\mathbb W}\right)-1\right)\right]\\
&=z\left(1+c+\mathbb E_{\mathbb Q_n}\left[\tilde\Xi-\log\left(\frac{d\mathbb Q_n}{d\mathbb W}\right)\right]\right)-z\log z=z.
\end{align*}
This together with (\ref{8}) gives (\ref{7}) and completes the proof.
\qed
${}$\\
${}$\\
We now have all the pieces in place that we need for the \textbf{completion of the proof of Theorem  \ref{thm1}}.
${}$\\
${}$\\
\textbf{Proof:}
The proof will be done in two steps. \\
${}$\\
\textbf{Step I:}
In this step we prove the existence and the uniqueness of the solution to (\ref{3}).
To this end, for $s\in [0,T]$ define
$F_s: L^2[s,\tau^{-1}(s)]\rightarrow L^2[s,\tau^{-1}(s)]$ by
$(F_s(\varphi))(t):=\int_{s}^{\tau^{-1}(s)} f(t,u) \varphi(u) du$, $\varphi\in L^2[s,\tau^{-1}(s)]$, $t\in [s,\tau^{-1}(s)]$.
Observe that (almost surely)
$F_s$ is a Hilbert-Schmidt operator.
From the fact that $\kappa(t,s)=0$ for $t<\tau^{-1}(s)$ we obtain that
for $t\in [s,\tau^{-1}(s)]$ (\ref{3}) is equivalent to the Fredholm integral equation
$(Id-F_s) \left(g(\cdot,s)\right)=-f(\cdot,s)$.
The relation $Spec(F)\subset (-\infty,1)$ implies that the map
$Id-F_s:L^2[s,\tau^{-1}(s)]\rightarrow L^2[s,\tau^{-1}(s)]$ is positive definite and so, there exists a unique solution
$g\in \mathcal V$
to the above Fredholm integral equation.
For $t\geq\tau^{-1}(s)$, equation
(\ref{3}) is equivalent to the Volterra equation (consider $t$ as a fixed parameter)
$$
f(t,s)-\kappa(t,s)=\int_{s}^{\tau^{-1}(s)} \left(f(t,u)-\kappa(t,u)\right) g(u,s) du, \ \ 0\leq s\leq\tau(t).
$$
Since $g\in L^2 ([0,T]^2)$ then from the standard theory of Volterra equations 
(see Section 2.7 in \cite{S:58}) the above equation has a unique solution. This completes the first step. \\
${}$\\
\textbf{Step II:}
In this step we complete the proof.
The uniqueness of the optimal trading strategy is due to strict concavity of the exponential utility.
Thus, in view of Lemma \ref{lem2} in order to complete the proof
it remains to establish (\ref{5}). First, the relation $\kappa(t,s)=0$ for $t<\tau^{-1}(s)$
implies that $\hat\gamma\in\mathcal A$.

As in the proof of Lemma \ref{lem2}, $g\in\mathcal V$ implies that there exists a probability measure
$\mathbb Q\sim\mathbb P$ and a
$\mathbb Q$-Brownian motion $(B_t)_{0\leq t\leq T}$
such that
$X_t=B_t-\int_{0}^t \int_{0}^s g(s,u)dB_u ds$, $t\in [0,T]$.
From Theorem 3 in \cite{C:03}
and (\ref{3})-(\ref{4}) we obtain the equality
$\frac{d\mathbb Q}{d\mathbb W}=e^{\hat c+\hat\Xi}$ where $\hat c\in\mathbb R$ is a constant and
$\hat\Xi:=\int_{0}^T\int_{0}^s f(s,u)dX_udX_s-\int_{0}^T\left(\hat\gamma_t-a(t)\right) dX_t$.
Clearly, $\hat c=\mathbb E_{\mathbb Q}\left[\log\left(\frac{d\mathbb Q}{d\mathbb W}\right)-\hat\Xi\right]$.
Thus, from (\ref{1}) and the relation $\mathbb E_{\mathbb W}\left[e^{\hat\Xi}\right]=e^{-\hat c}$
\begin{align}\label{9}
&\mathbb E_{\mathbb P}\left[\exp\left(-\int_{0}^T \hat\gamma_t dX_t\right)\right]=\mathbb E_{\mathbb W}\left[e^{c+\hat\Xi}\right]\nonumber\\
&=\exp\left(c-\mathbb E_{\mathbb Q}\left[\log\left(\frac{d\mathbb Q}{d\mathbb W}\right)-\hat\Xi\right]\right).
\end{align}

Next, similarly to (\ref{7+})
\begin{equation}\label{10}
\mathbb E_{\mathbb Q}\left[\log\left(\frac{d\mathbb Q}{d\mathbb W}\right)\right]=\frac{1}{2}\int_{0}^{T}\int_{0}^s g^2(s,u)du ds.
\end{equation}
Using the same arguments as in
(\ref{7++}) we obtain
\begin{align}\label{11}
&\mathbb E_{\mathbb Q}\left[\hat\Xi\right]=\mathbb E_{\mathbb Q}\left[\int_{0}^T\int_{0}^s \left(f(s,u)-\kappa(s,u)\right)dX_udX_s\right]\nonumber\\
&=-\int_{0}^T\int_{0}^s \left(f(s,u)-\kappa(s,u)\right))\tilde g(s,u)du ds.
\end{align}

Finally, from the properties $\kappa(t,s)=0$ for $t<\tau^{-1}(s)$ and $g(t,s)=0$ for $t\geq \tau^{-1}(s)$  we obtain
$\int_{0}^s \kappa(s,u)\tilde g(s,u)du=0$ for all $s$.
This together with (\ref{9})-(\ref{11}) yields (\ref{5}) and completes the proof.
\qed

\section*{Acknowledgments} This research was partially supported by the Israel Science Foundation grant 305/25.

\end{document}